\documentclass[pre,aps,twocolumn,superscriptaddress,amsmath,showpacs,floatfix]{revtex4}
\usepackage{graphicx}

\begin{document}
\title{Bidirectional cooperative motion of myosin-II motors on actin
tracks with randomly alternating polarities}
\author{Barak Gilboa\footnote{Authors with equal contribution}}
\affiliation{Department of Physics, Ben Gurion University,
Be'er Sheva 84105, Israel}
\affiliation{Department of Chemical Engineering, Ben Gurion University,
Be'er Sheva 84105, Israel}
\author{David Gillo$^*$}
\affiliation{Department of Chemical Engineering, Ben Gurion University,
Be'er Sheva 84105, Israel}
\author{Oded Farago}
\affiliation{Department of Biomedical Engineering, Ben Gurion University,
Be'er Sheva 84105, Israel}
\author{Anne Bernheim-Groswasser}
\affiliation{Department of Chemical Engineering, Ben Gurion University,
Be'er Sheva 84105, Israel}
\begin{abstract}
The cooperative action of many molecular motors is essential for
dynamic processes such as cell motility and mitosis. This action can
be studied by using motility assays in which the motion of
cytoskeletal filaments over a surface coated with motor proteins is
tracked. In previous studies of actin-myosin II systems, fast
directional motion was observed, reflecting the tendency of myosin II
motors to propagate unidirectionally along actin filaments. Here, we
present a motility assay with actin bundles consisting of short
filamentous segments with randomly alternating polarities. These actin
tracks exhibit bidirectional motion with macroscopically large time
intervals (of the order of several seconds) between direction
reversals. Analysis of this bidirectional motion reveals that the
characteristic reversal time, $\tau_{\rm rev}$, does not depend on the
size of the moving bundle or on the number of motors, $N$. This
observation contradicts previous theoretical calculations based on a
two-state ratchet model (Badoual et al., {\em
Proc.~Natl.~Acad.~Sci. USA}\/, 2002, {\bf 99}, 6696), predicting an
exponential increase of $\tau_{\rm rev}$ with $N$. We present a
modified version of this model that takes into account the elastic
energy due to the stretching of the actin track by the myosin II
motors. The new model yields a very good quantitative agreement with
the experimental results.
\end{abstract}

\maketitle

\section{Introduction}
\label{introduction}

Cells utilize biological motors for active transport of cargo along
their respective filaments to specific destinations
\cite{ref1}. Various types of motor proteins have different preferred
directions of motion. Most kinesins and myosins, for instance, move
towards the plus end of microtubules (MTs) and actin filaments,
respectively \cite{ref2}. Others, such as Ncd and myosin VI, move
towards the minus end \cite{ref3,ref4}. While some processes, such as
the transport of cargoes is achieved mainly by the action of
individual motors, other processes, such as cell motility and mitosis,
require the cooperative work of many motors. Muscle contraction, for
instance, involves the simultaneous action of hundreds of myosin II
motors pulling on attached actin filaments and causing them to slide
against each other \cite{ref5}. Similarly, groups of myosin II motors
are responsible for the contraction of the contractile ring during
cytokinesis \cite{ref6}. In certain biological systems, cooperative
behavior of molecular motors produces oscillatory motion. In some
insects, for instance, autonomous oscillations are generated within
the flight muscle \cite{ref7}. Spontaneous oscillations have also been
observed in single myofibrils in vitro \cite{ref8}. Finally, dynein
motors could be responsible for the oscillatory motion of axonemal
cilia and flagella \cite{ref9,ref10}.

The directionality of individual motors stems from interactions
between different parts of the motor and from interactions between the
motor and the track filament \cite{ref11,ref12}. The direction of
motion of a large collection of motors may also be influenced by their
cooperative mode of action. Specifically, in several recent
experiments the ability of motors to cooperatively induce
bidirectional motion has been demonstrated. These {\em in vitro}\/
experiments were performed by using motility assays in which a
filament glides over a dense bed of motors. In one such experiment,
unidirectional motion of actin filaments due to the action of myosin
II motors was transformed into bidirectional motion by the application
of an external stalling electric field \cite{ref13}. Under such
conditions the external forces acting on the actin filament nearly
balance the forces generated by the motors. Electric field was also
used to bias the direction of motion in kinesin-MT systems
\cite{ref13b}. In another experiment, bidirectional motion of MTs was
observed when subjected to the action of an ensemble of NK11 motors
\cite{ref14}. These motors are a mutant form of the kinesin related
Ncd, which individually exhibit random motion with no preferred
directionality \cite{ref14}. More recently, the motion of MTs on a bed
of a mixed population of plus-end (kinesin-5 KLP61F) and minus-end
(Ncd) driven motors was shown to exhibit dynamics whose directionality
depends on the ratio of the two motor species, including bidirectional
movement over a narrow range of relative concentrations around the
``balance point'' \cite{ref15}. Similarly, bidirectional transport of
microspheres coated with kinesin (plus-end directed) and dynein
(minus-end directed) on MTs was also reported \cite{ref15b}.

Several aspects of cooperativity in molecular motor systems have been
addressed using different theoretical models
\cite{ref15,ref16,ref17,ref18,ref19,ref20,ref21,ref22,ref23,ref24}. One
feature which has not been treated in these studies is the dependence
of the motion on the number of acting motors. A notable exception is
the work of Badoual {\em et al.}\/ \cite{ref18}, where a two-state
ratchet model has been used to examine the bidirectional motion
observed in the NK11-MT motility assay described in
ref.~\cite{ref14}. The model of Badoual {\em et al.}\/ \cite{ref18}
demonstrated the ability of a large group of motors working
cooperatively to induce bidirectional motion, even when individually
the motors do not show preferential directionality. (The model also
predicts that directional motors can also induce bidirectional
movement, if the filaments are close to stalling conditions in the
presence of an external load.) According to this model, the
characteristic time in which the filament undergoes direction reversal
(``reversal time''), $\tau_{\rm rev}$, increases exponentially with
the number of motors, $N$. Thus, the reversal time diverges in the
``thermodynamic limit'' $N\rightarrow\infty$, and the motion persists
in the direction chosen at random at the initial time.

In this work, we present an {\em in vitro}\/ motility assay in which
myosin II motors drive the motion of globally a-polar actin
bundles. These a-polar bundles are generated from severed (polar)
actin filaments whose fragments are randomly recombined. When
subjected to the action of a bed of myosin II motors, these a-polar
bundles exhibit bidirectional motion with characteristic reversal
times that are in the range of $\tau_{\rm rev}\sim 3-10$ sec. The
reversal times of the dynamics show no apparent correlation with the
size of the gliding bundles, or equivalently, with the number of
motors $N$ interacting with the track (which, because of the
homogeneous spreading of the motors on the bed, is expected to be
proportional to the size of the moving bundle). This observation is
clearly in disagreement with the strong exponential dependence of
$\tau_{\rm rev}$ on $N$, predicted by Badoual {\em et
al.}\/\cite{ref18}.

Here, we propose a modified version of this model that explains the
experimentally observed independence of $\tau_{\rm rev}$ on $N$. We
argue that the origin of this behavior can be attributed to the
tension developed in the actin track due to the action of the attached
myosin II motors. An increase in the number of attached motors leads
to an increase in the mechanical load which, in turn, leads to an
increase in the detachment rate of the motors, as already suggested in
models of muscle contraction \cite{ref21,ref22,ref23,ref24}. Unlike
most previous studies where the myosin conformational energy was
calculated, in this work we consider the elastic energy stored in the
actin track and demonstrate that the detachment rate increases
exponentially with $N$. This unexpectedly strong effect (which is
another, indirect, manifestation of cooperativity between the motors)
suppresses the exponential growth of $\tau_{\rm rev}$ with $N$.

\section{Materials and Methods}
\label{materialsandmethods}

\subsection{Protein purification} 
\vspace{-0.1cm}
Actin was purified from rabbit skeletal muscle acetone powder
\cite{ref25}. Purification of myosin II skeletal muscle is done
according to standard protocols \cite{ref26}. Actin labeled on Cys374
with Oregon Green (OG) purchased from Invitrogen.

\vspace{-0.3cm}
\subsection{NEM myosin II} 
\vspace{-0.1cm}
N-ethylmaleimide (Sigma, Co.) inactivated myosin II was prepared
according to standard protocol of Khun and Pollard~\cite{ref27}.

\vspace{-0.3cm}
\subsection{Optical Microscopy} 
\vspace{-0.1cm} Actin assembly was monitored for 30 minutes by
fluorescence with an Olympus IX-71 microscope. The labeled actin
fraction was $1/10$ and the temperature at which the experiments were
conducted was $23^{\circ}{\rm C}$. Time-laps images were acquired
using a DV-887 EMCCD camera (Andor Co., England).

\vspace{-0.3cm}
\subsection{Motility assay} 
\vspace{-0.1cm} Protocol for this assay was adopted from Kuhn {\em et
al.}\/ \cite{ref27}. The assay includes two essential steps: (a)
immobilization of actin filaments on a bed of NEM myosin II
inactivated motors, and (b) addition of active myosin II motors at a
defined concentration. For that purpose, 7.5-8.5$\mu$l of 0.2$\mu$M
NEM myosin II is introduced into a flow chamber (26mm$\times$2mm glass
surface area) for 1 minute of incubation followed by wash of the flow
chamber with BSA solution to passivate the surface. Following this,
actin filaments were grown on the surface (3$\mu$M 10\%
O.G. labeled). Finally, the cell was supplemented with 8$\mu$l of
0.6$\mu$M myosin II motors (in 2X myosin solution containing: 3.3mM
${\rm MgCl_2}$. 2mM EGTA, 20mM HEPES pH=7.6, 1\% MethylCellulose,
3.34mM Mg-ATP, 400mM DTT, 17.6mM Dabco), supplemented with 0.133M KCl,
5$\mu$M Vitamin D, and an ATP regenerating system containing 0.1mg/ml
Creatine Kinase (CK) and 1mM Creatine Phosphate (CP). At the KCl
concentrations used in this assay, the myosin II motors are assembled
in small motor aggregates ($\sim$16 myosin II units/aggregate) also
known as mini-filaments \cite{ref28}. Fluorescent images were taken
every 2 seconds for 30 minutes.

\vspace{-0.3cm}
\subsection{Data Analysis}
\vspace{-0.1cm}
The position of fluorescent bundles was determined as the intensity
center of mass using METAMORPH (Molecular Devices) software. The
position was analyzed using a custom MATLAB (The MathWorks, Inc.)
program. The data was corrected for stage drift. We first measured the
fluctuations of the positions of the bundles in the absence of ATP
(i.e., when the motors are not active). Under such conditions, the
positions of the bundles measured every 1 sec exhibit a Gaussian
distribution with zero mean and standard deviation $\Delta\sim 200$ nm
(Supplementary Information ? Figure S1). Bidirectional motion (with
ATP) was evaluated based on snapshot taken every 2 sec. We set
$\Delta$ as the experimental uncertainty since position changes that
are smaller then $\Delta$ cannot be unambiguously identified as a
change in the direction of motion caused by the action of the motors.

\vspace{-0.3cm}
\subsection{Estimation of number of interacting motors}
\vspace{-0.1cm} In order to evaluate the number of acting motors, the
dimensions of the bundles and the motors surface concentration, $C_m$,
must be determined. We estimate $C_m$ by assuming that all the motors
that were introduced into the flow chamber adhere to the top and
bottom glass surfaces of the flow cell (total surface, 104 ${\rm
mm}^2$). This gives $C_m\sim 27800\ (\mu m)^{-2}$, which corresponds
to densely packed motor beds (typical distance of a few nanometers
between motor heads). At such high densities, inhomogeneities
associated with the assembly of motors into mini-filaments can be
ignored.

The length of a bundle, $L$, was measured using METAMORPH
software. The width of a bundle was estimated by dividing its
fluorescence intensity by the intensity of single actin filaments,
which gives an estimate for the number of filaments, $N_f$, composing
the bundle. Assuming that the shape of the bundle is cylindrical, its
radius can be estimated as $R=\sqrt{N_f}\cdot r$, where $r\sim 3.75$
nm is the actin filament radius. The motors can only interact with the
part of the bundle that faces the myosin bed covered surface. Assuming
that this part corresponds to roughly a quarter of the surface of the
bundle, we find that the area that comes into contact with the motors
$A\sim(\pi/2)RL$. The number of interacting motors is, thus, given by
$N=C_mA$. Using this approximation and the measured surface
concentration and bundle dimensions, we estimate (see Fig.~\ref{fig6})
that $N\sim 1000-5000$ for most of the bundles studied in this work.

\vspace{0.3cm}
\subsection{Computer simulations}
\vspace{-0.1cm} A detailed discussion on the computational model is
found in the section \ref{discussion}, below. The model is based on
the model presented in ref.~\cite{ref18}, where $N$ rigidly coupled
equidistant motors interact with a one-dimensional periodic potential
representing the actin track. The spacing between the motors $q$ is
larger than and incommensurate with the periodicity of the potential,
$l$. The track consists of $M\simeq(q/l)N$ periodic units, which are
replicated periodically. In each unit of the track, a force of
magnitude $f_{\rm ran}$ and random directionality is introduced which
defines the local polarity of the track. Globally a-polar tracks were
generated by setting the total random force to zero (i.e., choosing an
equal number of periodic units in which the random forces point to the
right and left). The motion of the motors on these tracks was
calculated by numerically integrating the equations of motions
[$dx=(F_{\rm tot}/\lambda)dt$] (see Eq.~\ref{eq1} and following text)
with time step $dt=0.05$ msec.  The position of one of the motors
along the track was recorded every 0.25 sec, and changes in the
direction of the motion of the motors were identified by analyzing the
position of this motor. The distribution of reversal times, $t$,
follows an exponential distribution: $p(t)=(1/\tau_{\rm
rev})\exp(-t/\tau_{\rm rev})$, from which $\tau_{\rm rev}$ was
extracted. The error bars in Fig.~\ref{fig7}C represent one standard
deviation of the distribution of reversal times measured for different
realizations of globally a-polar tracks of similar size. For each
value of $N$, the number of simulated realizations is 40.

\section{Results}
\label{results}

\begin{figure}[t]
\begin{center}
\scalebox{0.9}{\centering \includegraphics{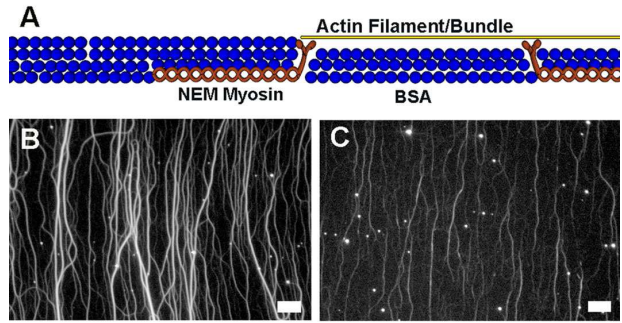}}
\end{center}
\vspace{-0.5cm}
\caption{(A) Schematic diagram of the system before addition of active
motors. The surface of a microscope slide was saturated by BSA (blue
balls) and NEM myosin II (long, two-headed, brown objects). Actin
filaments/bundles (thin yellow line) are attached to NEM myosin heads
above the surface. (B,C) Images of the system before the addition of
active myosin II minifilaments. (B)\ shows the thick actin bundles
formed at a high concentration of ${\rm MgCl_2}$(1.67 mM), while in
(C) the thin bundles/filaments formed at a low ${\rm MgCl_2}$
concentration (0.5 mM) are shown. Bar size is 5 $\mu$m. }
\label{fig1}
\end{figure}

A diagrammatic representation of the system is displayed in
Fig.~\ref{fig1}. The protocol, based on that of Kuhn {\em et
al.}\/ \cite{ref27}, is described in detail above. In brief, the
surface of a microscope slide was saturated with NEM-inactivated
myosin II motors (drawn as long, two headed, brown objects at the
bottom of Fig.~\ref{fig1}A) and passivated by BSA (blue balls in
Fig.~\ref{fig1}A). Subsequently, actin filaments/bundles (thin yellow
line, Fig.~\ref{fig1}A) were grown and held firmly on the underside of
the NEM-myosin II bed. Fig.~\ref{fig1}B shows a characteristic
fluorescent microscope image of the system which, at this stage,
consisted of a large number of long actin bundles. The bundles were
formed due to the presence of free ${\rm Mg^{2+}}$ ions (concentration
1.67 mM), which induced attractive electrostatic interactions between
the actin filaments \cite{ref29}. Unlike bundles formed by certain
actin-binding proteins, filaments formed by condensation in the
presence of multivalent cations are randomly arranged within the
bundles without any specific polarity \cite{ref30,ref31}. At lower
concentrations of ${\rm Mg^{2+}}$ (0.5?mM), both thinner bundles
and single filaments were observed (Fig.~\ref{fig1}C).

\begin{figure}[t]
\begin{center}
\scalebox{0.9}{\centering \includegraphics{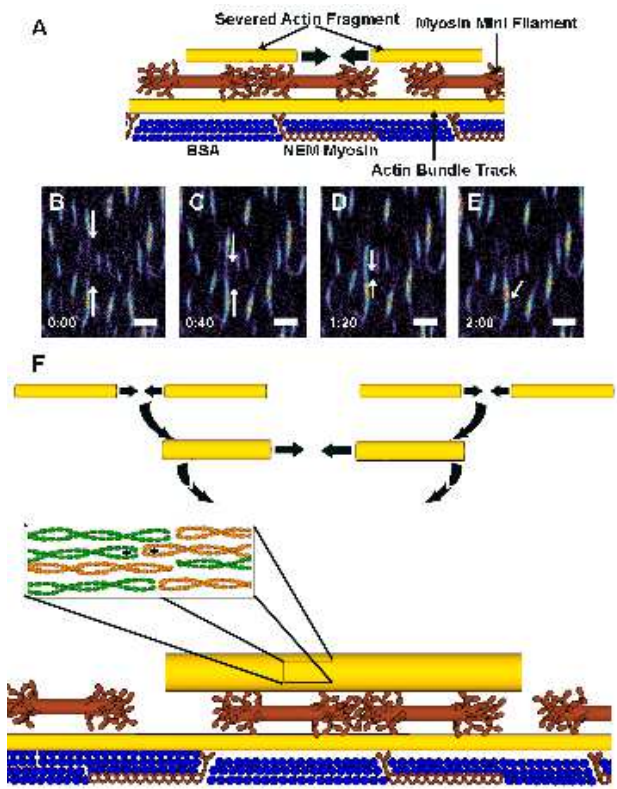}}
\end{center}
\vspace{-0.5cm}
\caption{(A) Schematic diagram of the system after addition of active
myosin II motors. After the initial step {\protect (see
Fig.~\ref{fig1})}, myosin II minifilaments (multi-headed brown
objects) were added to the cell sample. The motors that landed on the
BSA surface created a homogeneous bed of immobile, yet active,
motors. Other motors landed on the actin filaments/bundles (long
yellow line) present on the surface. The myosin II minifilaments
started to move along the actin filaments/bundles. During their
motion, the motors exerted forces on the actin filaments, which caused
severing of small actin fragments (short yellow lines). The ruptured
actin fragments could move rapidly on the bed of active myosin II
minifilaments and fuse with other bundles. One fusion event is
demonstrated in the sequence of snapshots (B-E). Here, we show (B) two
bundles moving oppositely to each other, getting closer (C) and then
fusing (D-E) to create one larger object. Time is given in minutes,
bar size is 5 $\mu$m. (F) The bundles continue to grow in size through
multiple fusion processes, until eventually a large, highly a-polar
bundle is formed (thick yellow tube - the inset illustrates the
internal structure of such a bundle, consisting of individual actin
filaments with randomly orientated polarities).}
\label{fig2}
\end{figure}

\begin{figure}[t]
\begin{center}
\scalebox{0.65}{\centering \includegraphics{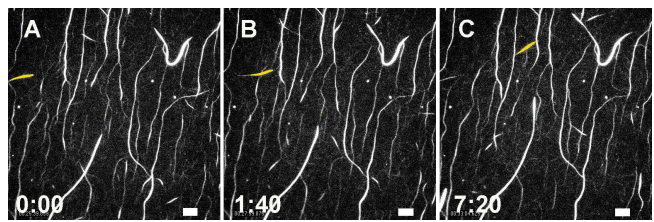}}
\end{center}
\vspace{-0.5cm}
\caption{Sequence of snapshots showing actin bundles moving
directionally on a bed of active myosin II motors, at a low motor
concentration ([myosin II motor]= 0.3 $\mu$M). As a visual guide, we
have marked one such actin bundle in orange. This bundle moved
directionally across the image plane from left to right at times (A)
0:00, (B) 1:40 min, and (C) 7:20 min (typically at a velocity of 10
$\mu$m/min). Bar size is 5 $\mu$m.}
\label{fig3}
\end{figure}

After the initial step, myosin II minifilaments (multi-headed brown
objects, Fig.~\ref{fig2}A) were added to the cell sample. The motors
that landed on the BSA surface created a homogeneous bed of immobile,
yet active, motors. Other motors landed on the actin filaments/bundles
present on the surface. These motors started to move along the actin
tracks, thereby exerting forces on the actin filaments, which led to
the severing of small actin fragments \cite{ref32,ref33}
(Fig.~\ref{fig2}A). The ruptured actin fragments were then free to
move rapidly on the bed of active myosin II motors. When gliding
bundle fragments came into close proximity to other bundles, they
could fuse, creating new, a-polar, bundles (Fig.~\ref{fig2}A. See also
Supplementary Information - movie 1, showing a small actin piece
severed from a filament, moving rapidly and fusing with a distant
existing bundle). These newly created bundles could further fuse with
each other to form even larger objects (see Fig.~\ref{fig2}F and the
sequence of snapshots in Figs.~\ref{fig2}B-E depicting one such event
of fusion of bundles). The rate of fusion events decreased with time
and, after several minutes, the system relaxed into its final
configuration, shown schematically in Fig.~\ref{fig2}F). Notice that
the severing and rearrangement of the originally formed actin
filaments/bundles (Fig.~\ref{fig1}B) led to the formation of much
shorter bundles (Figs.~\ref{fig2}B-E). Moreover, the random nature of
the multiple fusion processes involved in the generation of these
shorter bundles ensured that the final actin tracks were highly
a-polar. Indeed, the motion of most of the bundles shown in
Supplementary Information - movie 2 was bidirectional (``back and
forth'' motion), and only those bundles undergoing rare fusion events
exhibited unidirectional motion. It is important to emphasize that
bidirectional motion was observed only above a certain concentration
of added myosin II motors (0.6 $\mu$M) and only in the presence of
ATP. At lower concentrations of motors (0.3 $\mu$M), the motion of
actin bundles was directional (see Fig.~\ref{fig3} and Supplementary
Information - movie 3. Note that in movie 3, the motion takes place
both along pre-existing actin tracks, as well as on the BSA bed, both
which are covered by active myosin II mini-filaments. The motion
between these two areas is continuous, demonstrating that the whole
surface is covered uniformly with motors.) We, therefore, conclude
that the bidirectional movement originates from the action of the
active myosin II motors which (i) severed actin pieces, (ii)
transported the severed fragments, which fused into actin tracks with
randomly alternating polarities, and (iii) moved these a-polar actin
tracks bidirectionally (see Supplementary Information - movie 2).

\begin{figure}[t]
\begin{center}
\scalebox{0.725}{\centering \includegraphics{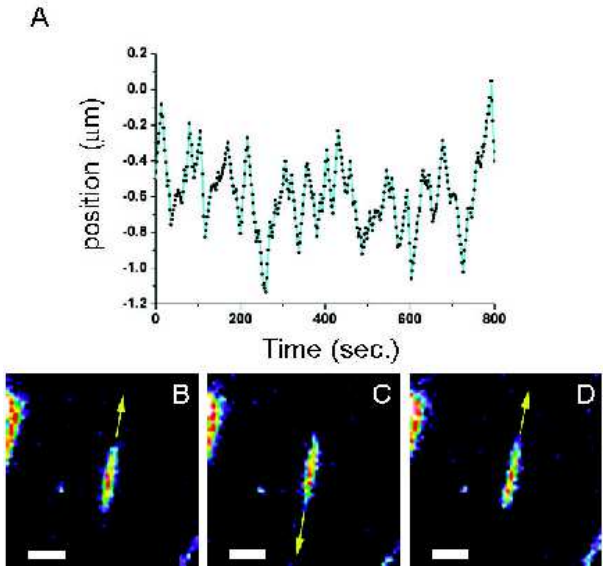}}
\end{center}
\vspace{-0.5cm}
\caption{Position of a bundle over a time interval of 800 sec. The
time interval between the consecutive data points is 2 sec. (B-D)
Pseudo-color images of the actin bundle. The yellow arrows indicate
the instantaneous direction of motion of the bundle. Bar size is 5
$\mu$m.}
\label{fig4}
\end{figure}

\begin{figure}[h]
\begin{center}
\scalebox{1.1}{\centering \includegraphics{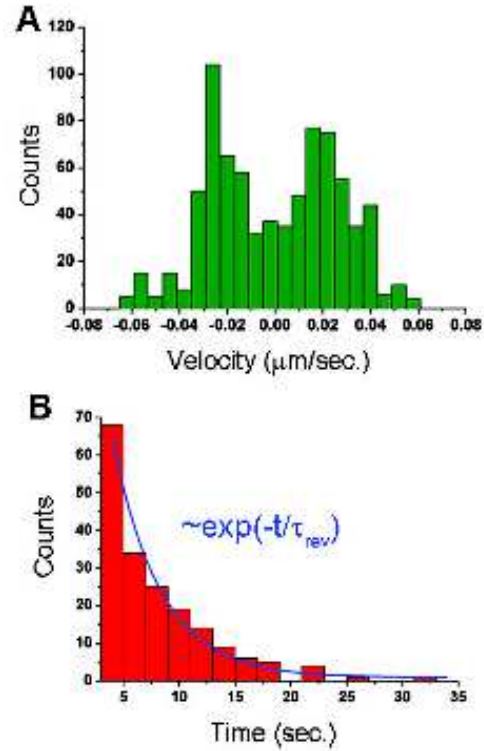}}
\end{center}
\vspace{-0.5cm}
\caption{ Velocity histogram of the bundle whose motion is shown in
{\protect Fig.~\ref{fig4}} (based on 900 sampled data points ?
Figure {\protect \ref{fig4}} shows only 400 of those points),
exhibiting a clear bimodal distribution. (B) Distribution of the
reversal time for the same bundle. The distribution is fitted by a
single exponential decay function with a characteristic reversal time:
$\tau_{\rm rev}\sim 3$ sec.}
\label{fig5}
\end{figure}

\begin{figure}[t]
\begin{center}
\scalebox{0.9}{\centering \includegraphics{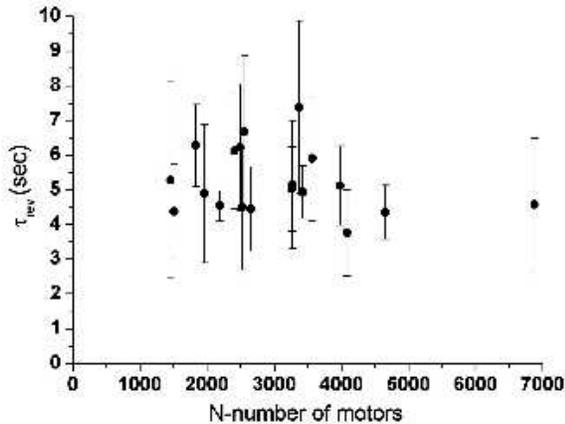}}
\end{center}
\vspace{-0.5cm}
\caption{Characteristic reversal time, $\tau_{\rm rev}$, of 19
different bundles as a function of the number of working motors
$N$. The reversal time for each bundle is obtained by an exponential
fit.}
\label{fig6}
\end{figure}

Fluorescence microscopy was used to follow the bidirectional motion of
the actin tracks. Fig.~\ref{fig4}A shows the position of center of
mass of one bundle (three snapshots are shown in Figs.~\ref{fig4}B-D)
during a period of more than 10 minutes of the experiment. The
dynamics of this bundle are representative of the motion of the other
actin bundles. Specifically, the one-dimensional motion of the bundle
does not persist in the initial direction, but rather exhibits
frequent direction changes. Measurements of the position of the center
of mass of the bundle were taken at time intervals of $\Delta t=2$
sec, and the mean velocity in each such period of motion was evaluated
by $v=\Delta x/\Delta t$, where $\Delta x$ is the displacement of the
center of mass (see section
\ref{materialsandmethods}). Fig.~\ref{fig5}A shows the velocity
histogram of the bundle shown in Fig.~\ref{fig4}. The velocity
histogram is bimodal indicating bidirectional motion. The speed of the
bundle varies between $|v|=1-2$ $\mu$m/min, which is 2 orders of
magnitude lower than the velocities measured in gliding assays of
polar actin filaments on myosin II motors \cite{ref34}. The fact that
the typical speed of the bidirectional motion is considerably smaller
than those of directionally-moving polar actin filaments can be
partially attributed to the action of individual motors working
against each other in opposite directions. The bidirectional movement
consists of segments of directional motion which typically last
between 2 to 10 time intervals of $\Delta t=2$. The statistics of
direction changes is summarized in Fig.~\ref{fig5}B which shows a
histogram of the number of events of directional movement of duration
$t$. The characteristic reversal time, $\tau_{\rm rev}$, can be
extracted from the histogram by a fit to an exponential distribution:
$p(t)=(1/\tau_{\rm rev})\exp(-t/\tau_{\rm rev})$. This form (which, as
exemplified in Fig.~\ref{fig5}B, fits the data well) is expected if
the probability per unit time to "turn" in the opposite direction is
independent of the time since the beginning of motion in a given
direction.

Although Figs.~\ref{fig4} and \ref{fig5} summarize the results
corresponding to the movement of a single actin bundle, these results
are representative of several tens of bundles whose motion we followed
in several repeated experiments. In these experiments we observed that
essentially {\em all}\/ a-polar bundles exhibited bidirectional motion. For
the sake of our quantitative analysis, we picked a smaller group of 19
bundles (about 25\% of all bundels) for which both the reversal time
and the number of acting motors $N$ (which is proportional to the
surface area, see section \ref{materialsandmethods}) could be
determined with sufficiently high precision.  The results
corresponding to the motion of this sample of representative bundles
are plotted in Fig.~\ref{fig6}. The choice of which bundles to include
in Fig.~\ref{fig6} is based on the following practical reasons: The
fluoresces intensity of small bundles is too low (compared with the
background) and, thus, the accurate position and dimensions are hard
to determine. Very large bundles are practically immobile and their
motion is smaller than the experimental spatial resolution (see
Material and Methods - section \ref{materialsandmethods}). Thus, the
data in Figure \ref{fig6} includes only bundles of ``intermediate''
size. One can see in Fig.~\ref{fig6} that while $N$ varies over half
an order of magnitude, the corresponding $\tau_{\rm rev}$ are similar
to each other ($3<\tau_{\rm rev}<10$ sec) and show no apparent
correlation with $N$.

\begin{figure*}[t]
\begin{center}
\scalebox{1}{\centering \includegraphics{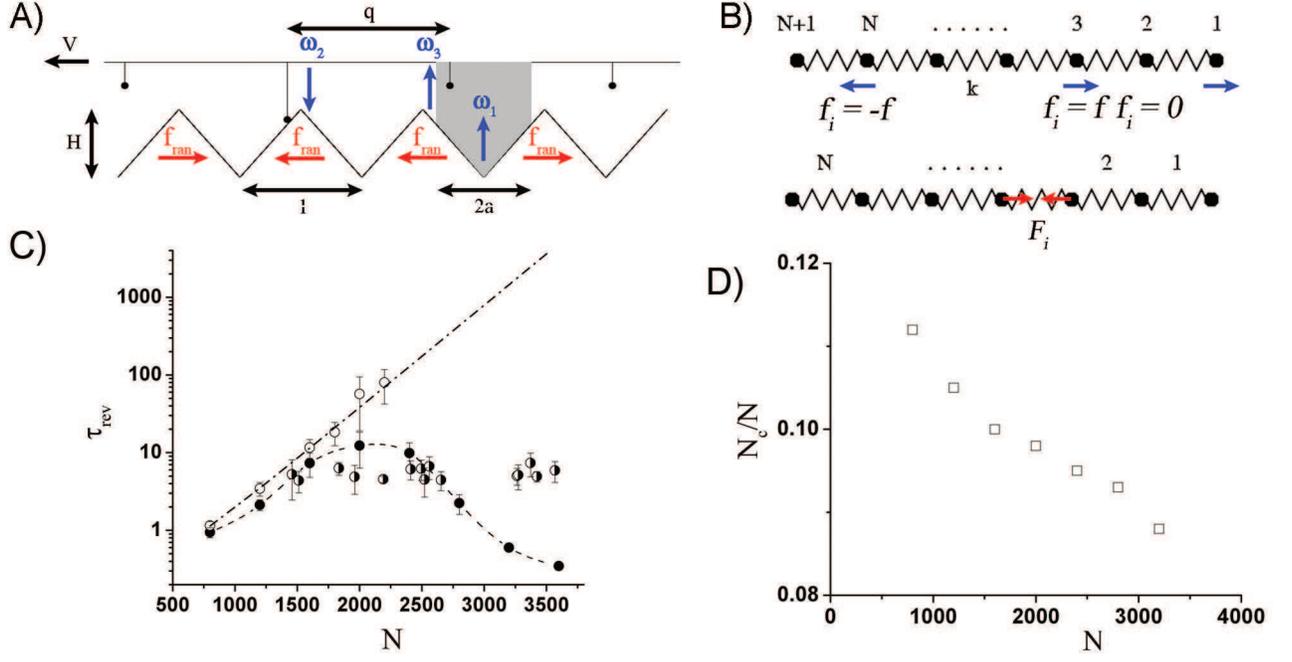}}
\end{center}
\vspace{-0.5cm}
\caption{(A) $N$ point particles (representing the motors) are
connected to a rigid rod with equal spacing $q$. The motors interact
with the actin track via a periodic, symmetric, saw-tooth potential
with period $l$ and height $H$. In each periodic unit, there is a
random force of size $f_{\rm ran}$, pointing either to the right or to
the left (red arrows). The motors are subject to these forces only if
connected to the track. The detachment rate $\omega_1$ is localized in
the shaded area of length $2a<l$, while the attachment rate $\omega_2$
is located outside of this region. The off rate $\omega_3$ is
permitted only outside the gray shaded area. (B) A set of $n+1$ point
particles connected via $N$ identical springs with spring constant
$k$. Each particle is subjected to a random force $f_i$ that takes
three possible values: $-f,\ 0,\ +f$ (blue arrows). The mean force
acting on the particles is given by $\bar{f}=f_{\rm CM}/(n+1)$, where
$f_{\rm CM}=\sum_{i=1}^{N+1}f_i$ is the total force acting on the
center of mass and causing the movement of the system. The force $F_i$
stretching the $i$-th spring (red arrow) is equal to the sum of the
excess forces, $f_i-\bar{f}$, acting on the particles located to the
right of the $i$-th spring,
$F_i=(\sum_{j=1}^{i}f_j)-i\bar{f}$. Because a similar expression can
be written taking into account the forces acting on the particles
located to the left of the spring, one can readily show that
$\sum_{i=1}^{N}F_i=0$. (C) The mean reversal time, $\tau_{\rm rev}$,
as a function of the number of motors $N$, computed for different
realizations with $\alpha=0.0018$ (solid circles and a dashed line as
a guide to the eye) and $\alpha=0$ (open circles). In the latter case,
the effect of actin stretching is neglected, and $\tau_{\rm rev}$
grows exponentially with $N$ (dash-dotted line), in agreement with
{\protect ref.~\cite{ref18}}. The half-filled circles denote the
experimental results, also presented in Fig.~{\protect
\ref{fig6}}. (D)\ The fraction of attached motors, $N_c/N$, as a
function of the total number of motors, for $\alpha=0.0018$. The
values of $N_c/N$ for $\alpha=0$ are indistinguishable.}
\label{fig7}
\end{figure*}

\section{Discussion}
\label{discussion}

The bidirectional motion of motors was previously seen in systems
consisting of motors that lack specific directionality
\cite{ref14,ref35}, mixtures of motors working in opposite directions
\cite{ref15}, or under the action of external forces close to stalling
conditions (forces acting on the filament that nearly balance the
forces generated by the motors) \cite{ref13}.  One commonly used model
for the dynamics of molecular motors in the biophysics literature is
the Brownian ratchet mechanism \cite{ref36}. Within this modeling
approach, the motion of individual motor proteins is studied by
considering the motion of a particle in a periodic, locally
asymmetric, potential. It follows from the second law of
thermodynamics that if the system is coupled to a thermal bath, the
particle subjected to the periodic potential will not exhibit large
scale directed motion. Directed motion is possible only if the system
is (i) locally asymmetric, and (ii) driven out of equilibrium by an
additional deterministic or stochastic perturbation. This perturbation
is used in the model to represent the consumption of ATP chemical
energy by the motors. Ratchet models are not molecular in nature but
rather present a way to identify the minimal physical requirements for
the motion of motor proteins. However, by choosing properly the
parameters of system, they may be employed to derive quantitative
predictions for specific motor-filaments systems. Ratchet models have
been extended for describing and analyzing the collective motion of
groups of motors.  The motion of several motors is influenced by the
motor-motor interactions \cite{ref36b} and mechanical coupling
\cite{ref18}. The model proposed by Badoual {\em at al.}\/
\cite{ref18} (and which, below, we present a slightly modified version
of) demonstrates that mechanical coupling between the motors is
sufficient for the generation of highly cooperative bidirectional
motion, even if the motors attach to/detach from the track in an
uncorrelated fashion.

A key prediction of the model in Ref.~\cite{ref18} is the exponential
increase of the mean reversal time of the bidirectional motion,
$\tau_{\rm rev}$, with $N$, the number of motors. This prediction is
in contradiction with our experimental results
(Fig.~\ref{fig6}). Here, we show that this disagreement can be
resolved by considering the stretching energy involved in the
interactions between the actin track and the ``walking''
motors. Accounting for this effect eliminates the exponential
dependence of $\tau_{\rm rev}$ on $N$. Moreover, when values
representing myosin II-actin systems were assigned to the parameters
of the model, we found $\tau_{\rm rev}\sim 1-12$ sec, which in a very
good quantitative agreement with the experimental data.

Our model is illustrated schematically in Fig.~\ref{fig7}A: We
consider the 1D motion of a group of $N$ point particles (representing
the motors) connected to a rigid rod with equal spacing $q$. The actin
track is represented by a periodic saw-tooth potential, $U(x)$, with
period $l$ and height $H$. We choose $q=(5\pi/12)l\sim 1.309 l$, which
satisfies the requirements of the model \cite{ref18} for $q$ to be
larger than and incommensurate with the periodicity of the
potential. The locally preferred directionality of the myosin II
motors along the actin track is introduced via an additional force of
size $f_{\rm ran}$ exerted on the individual motors. In each unit of
the periodic potential, this force randomly points to the right or to
the left (the total sum of these forces vanishes), which mimics the
random, overall a-polar, nature of the actin bundles in our
experiments.

The instantaneous force between the track and the motors is given by
the sum of all the forces acting on the individual motors:
\begin{widetext}
\begin{equation}
F_{\rm tot}=\sum_{i=1}^{N}f_i^{\rm motor}=\sum_{i=1}^{N}\left[
-\frac{\partial U\left(x_1+\left(i-1\right)q\right)}{\partial x}+
f_{\rm ran}\left(x_1+\left(i-1\right)q\right)\right]\cdot C_i(t),
\label{eq1}
\end{equation}
\end{widetext}
where $x_i=x_1+(i-1)q$ is the coordinate of the $i$-th motor. The two
terms in the square brackets represent the forces due to the symmetric
saw-tooth potential and the additional random local forces acting in
each periodic unit. The latter are denoted by red arrows in
Fig.~\ref{fig7}A. The function $C_i(t)$ takes two possible values, 0
or 1, depending on whether the motor $i$ is detached or attached to
the track, respectively, at time $t$. The group velocity of the motors
(relative to the track) is determined by the equation of motion for
overdamped dynamics: $v(t)=F_{\rm tot}(t)/\lambda$. The friction
coefficient, $\lambda$, depends mainly on motors attached to the track
at a certain moment and is therefore proportional to the number of
connected motors, $N_c\leq N$ at time $t$: $\lambda=\lambda_0 N_c$.

To complete the dynamic equations of the model, we need to specify the
transition rates between states (0 - detached; 1 - attached). The
motors change their states independently of each other. We define an
interval of size $2a<l$ centered around the potential minima (the gray
shaded area in Fig.~\ref{fig7}A). If located in one of these regions,
an attached motor may become detached ($1\rightarrow 0$) with a
probability per unit time $\omega_1$. Conversely, a detached motor may
attach to the track ($0\rightarrow 1$) with transition rate $\omega_2$
only if located outside this region of size $2a$. However, we also
allow another independent route for the detachments of motors, which
may take place outside the gray shaded area in Fig.\ref{fig7}A (i.e.,
around the potential maxima) and is characterized by an off rate
$\omega_3$. The rates $\omega_1$, $\omega_2$, $\omega_3$ (see blue
arrows in Fig. 7A), represent the probabilities per unit time of a
motor to (i) detach after completing a unit step, (ii) attach to the
track, or (iii) detach from the track without completing the step.

Generally speaking, the rates of transitions between states depend on
many biochemical parameters, most notably the types of motors and
tracks, and the concentration of chemical fuel (e.g., ATP). They may
also be affected by the forces induced between the motors and the
filament, which result in increase in the configurational energy of
the attached myosin motors \cite{ref21,ref22,ref23,ref24} and in the
elastic energy stored in the S2 domains of the mini-filaments, as well
as an increase in the stretching energy of the actin filament. The
latter contribution can be introduced into the model via a modified
detachment rate given by: $\omega_3=\omega_3^0\exp(-\Delta E/k_BT)$,
where $\Delta E$ is the change in the elastic energy of the actin
track due to the detachment of one motor head. The dependence of
$\Delta E$ on the number of connected motors $N_c$ (out of a total
number of motors, $N$) can be estimated in the following manner:
Consider a series of $N+1$ point particles connected by $N$ identical
springs (representing a series of sections of actin filaments) having
a spring constant $k$ (see Fig. 7B). Let us assume that random forces
act on the particles and denote the force applied on the particle with
index i ($1\leq i\leq N+1$) by $f_i$. Assume that each of these forces
can take three possible values: $-f$ (representing attached motors
locally pulling the track to the left), $+f$ (attached motors pulling
the track to the right), and 0 (detached motors not applying
force). Defining the "excess force" with respect to the mean force
acting on the particles: $f_i^*=f_i-\bar{f}$ [where
$\bar{f}=\sum_{i=1}^{N+1}f_i/(N+1)$], one can show that the
force stretching (or compressing) the $i$-th spring in the chain is
given by the sum of excess forces acting on all the particles located
on one side of the spring
\begin{equation}\
F_i=\sum_{j=1}^i f_j^*=-\sum_{j=i+1}^{N+1}f_j^* .
\label{eq2}
\end{equation}
From Eq.~\ref{eq2} it can be easily verified that $\sum_{i=1}^N
F_i=0$. We thus conclude that the excess forces acting on the
particles, $f_i^*$, represent a series of random quantities with zero
mean. Therefore, the size of $F_i$ can be estimated by mapping the
chain of springs into the problem of a 1D random polymer ring
\cite{ref37}, where the elastic energy stored in the $i$-th spring,
$\epsilon_i=F_i^2/2k$, plays the role of the squared end-to-end
distance between the $i+1$ monomer and the origin. From this mapping
we readily conclude that the energy of most of the springs (except for
those located close to the ends of the chain) scales linearly with the
number of attached motors: $\epsilon\sim N_c(f^2/2k)$. The total
elastic energy of the chain scales as
\begin{equation}
E\sim N\epsilon\sim NN_c(f^2/2k),
\label{eq3}
\end{equation}
and when a motor detaches from the track ($N_c\rightarrow N_c-1$), 
\begin{equation}
\Delta E/k_BT=-\alpha N
\label{eq4}
\end{equation}
where $\alpha$ is a dimensionless prefactor.

We simulated the dynamics of an $N$-motor system, choosing parameters
corresponding to the myosin II-actin system. The period of the
potential $l=5$ nm corresponds to the distance between binding sites
along the actin track \cite{ref38,ref39,ref40}, and the amplitude of
the symmetric potential is set to $H=6 k_BT$. Thus, the force
generated by each motor head on the track is $2H/l=10$ pN (first term
in square brackets in Eq.~\ref{eq1}). The magnitude of the random
force that defines the local polarity of the track (second term) is
given by $f_{\rm ran}=4.5$ pN, so the total force acting on each motor
head ranges between about 5 to 15 pN
\cite{ref38,ref39,ref41,ref42}. The interval around the potential
minima from which motors can detach from the track with rate
$\omega_1$ is chosen to be $2a=3.8$ nm. The transitions rates between
attached and detached states are $\omega_1^{-1}=0.5$ ms and
$\omega_2^{-1}=33 $ ms \cite{ref43,ref44,ref45,ref46}. With this
choice of parameters, we obtain a system with a low fraction of
attached motors $Nc/N\sim0.1$ (see Fig.~\ref{fig7}D). We also set the
friction coefficient per attached motor to $\lambda_0=1.25\cdot 10^{-4}$
kg/s, which yields the experimentally measured velocity $v \sim0.03$
nm/ms $\sim2$ $\mu$m/min. The rate $\omega_3^0$ expresses the
probability of a single motor head to detach from the track without
advancing to the next unit. The probability $p$ of such an event is
1-2 orders of magnitude smaller than the complementary probability
$(1-p)$ to execute the step. We take $p = 1/30$ \cite{ref46}, which
yields ?$(\omega_3^0)^{-1}\sim pv/l=7500$ ms. Finally, the exponent
$\alpha$ appearing in Eq.~\ref{eq4} is evaluated by: 
\begin{equation}
\alpha\sim(f^2/2kk_BT)=(f^2l/2YAk_BT),
\label{eq5}
\end{equation}
where $Y\sim10^9$ Pa is Young's modulus for actin and $A\sim 35\ {\rm
nm}^2$ is the cross sectional area of an actin filament
\cite{ref46}. For the model parameters: $f\sim 10$ pN, $l=5$ nm, we
find $\alpha\sim 0.0018$.

Fig.~\ref{fig7}C shows the computationally measured reversal time
$\tau_{\rm rev}$ as a function of $N$ for $800\leq N\leq3600$. This
range largely overlaps with the estimated range of number of motors in
our experiments (see Fig.~\ref{fig6}. The experimental data points in
this range of $N$ are replotted in Fig.~\ref{fig7}C and denoted in
half-filled circles). For each $N$, the computational results
represent the average $\tau_{\rm rev}$ computed for 40 different
realizations of random, overall a-polar, tracks. The error bars
represent the standard deviation of $\tau_{\rm rev}$ between
realizations, where for each realization $\tau_{\rm rev}$ is estimated
by fitting the histogram of turning times to an exponential decay
function (as in the experimental part - see Fig.~\ref{fig5}B).  Two
sets of computational data are shown in Fig.~\ref{fig7}C: one
corresponding to $\alpha= 0.0018$ (solid circles), and the other to
$\alpha=0$ (open circles), i.e., without considering the effect of
actin stretching, but when all the other system parameters mentioned
above are kept unchanged. The latter case is qualitatively similar to
the model presented in ref.~\cite{ref18}, exhibiting a very strong
exponential dependence of $\tau_{\rm rev}$ on $N$ (indicated by the
straight dashed-dotted line in Fig.~\ref{fig7}C).  In contrast, the
data corresponding to $\alpha=0.0018$ show much weaker variation in
$\tau_{\rm rev}$ upon changing $N$. The mean reversal times computed
for $1200\leq N\leq 2800$ are found in the range $2\leq \tau_{\rm
rev}\leq 12$ sec, which is in a very good quantitative agreement with
the corresponding range of experimental results, and certainly does
not grow to values of thousands of seconds as predicted for
$\alpha=0$.  The validity of our ratchet model is quite remarkable in
view of its extreme simplicity; but one must be aware of the following
points of disagreement between the experimental and computational results
(which illustrate the limitations of the model): (1) The computed
reversal times show weak, non-monotonic, dependence on $N$ which is
not observed experimentally. (2) The largest computed $\tau_{\rm rev}$
($\tau_{\rm rev}=12$ sec for $N=2000$) is slightly larger than the
experimentally measured reversal times. (3) The computational results
for $N<1000$ and $N>3000$ cannot be directly compared with
experimental results since the corresponding reversal times
($\tau_{\rm rev}<1$ sec) fall below the experimental resolution.

The decrease of the computed reversal times for $N>2400$ can be
attributed to the ``mean field'' nature of calculation of $\omega_3$,
i.e., to our assumption that (for a given $N$) the detachment of each
motor head leads to the same energy gain (see Eq.~\ref{eq4}).  In
reality, the energy change upon detachment of a motor depends, in some
complex manner, on a number of factors such as the positions and
chemical states of the motors. Motors which release higher energy will
detach at higher rates, and the detachment of these ``energetic''
motors will lead to the release of much of the elastic energy stored
in the actin track. We, therefore, conclude that within the mean field
approach, the number of disconnecting motors and the frequency of
detachment events are probably over-estimated.  This systematic error
of the mean field calculation increases with $N$, and the result of
this is the decrease of $\tau_{\rm rev}$ in this regime, which is not
observed experimentally. For even larger values of $N$ ($N>5000$), the
model fails because $\omega_3>\omega_2$ and the effective attachment
rate outside the gray shaded area in Fig.~\ref{fig7}A, $\omega_{\rm
on}\equiv\omega_2-\omega_3$, becomes negative, i.e., motors detach
from the track faster than they attach to it. In contrast, for
$<1000<N<3000$, $\omega_3^0\ll \omega_3\ll\omega_2$, and the effective
attachment rate $\omega_{\rm on}$ barely changes upon changing the
model parameter from $\alpha=0$ (ref.~\cite{ref18}) to $\alpha=0.0018$
(our model). This seemingly minute change in $\omega_{\rm on}$ (which,
nevertheless, involves a dramatic increase in $\omega_3$) leads to the
following non-trivial outcome: On the one hand, the fraction of
attached motors remains unchanged. The data shown in Fig.~\ref{fig7}D
corresponds to {\em both}\/ values of $\alpha$ for which the results
for $N_c/N$ are indistinguishable. On the other hand, the reversal
times drop by as much as three orders of magnitude (for $N=3000$) when
$\alpha$ is modified from 0 to 0.0018.  This spectacular decrease in
$\tau_{\rm rev}$ is, therefore, not the result in the change in the
number of attached motors (since, for each value of $N$, the same
fraction of motors is attached for both values of $\alpha$), but
rather can be related to the less regular manner by which the motors
detach from the track. The more frequent stochastic detachments of
motors from the actin track increases the probability per unit time of
motion reversal.

\section{Conclusion}
\label{conclusion}

We have investigated the dynamics of myosin II motors on actin tracks
composed of small filamentous segments with randomly alternating
polarities. The absence of global polarity leads to a bidirectional
relative motion between the motors and the tracks. The characteristic
reversal time of this motion is of the order of a few seconds and
exhibits no particular dependence on the number of acting
motors. Bidirectional motion with macroscopic reversal times has been
previously observed for NK11 motors on microtubules and has been
attributed to the cooperativity of the motors. According to previously
proposed models, the signature of such a bidirectional cooperative
motion is the strong exponential dependence of $\tau_{\rm rev}$ on
$N$. The contradiction of this prediction with our experimental
results can be reconciled by incorporating an additional feature into
the model, namely, the effect of actin stretching by the walking
motors.  To reduce the associated elastic energy, the off rate of
motors increases, and many of them detach from the track before
completing a unit step.  This effect reduces $\tau_{\rm rev}$
considerably and eliminates its exponential growth with $N$.  

Single molecule experiments have led to a dramatic increase in our
understanding of the structure and dynamics of individual molecular
motors. However, many biological processes such as muscle contraction,
cytokinesis, and the motion of axonemal cilia and flagella, involve
cooperative action of many motors, which may be affected by the
structure of the underlying track.  This concept is clearly
demonstrated in this work dealing with the bidirectional motion of
myosin II motors on actin tracks with randomly alternating local
polarities, but without a net preferred directionality at the
mesoscopic level. This unique type of motion is induced by the forces
of individual motors whose collective effect is manifested in
macroscopically large reversal times.  At the same time, the
cooperativity of these forces also increases the elastic energy of the
track, and thereby limits the growth of $\tau_{\rm rev}$.

\section{Acknowledgment}
We thank Nir Gov, Yoav Tsori and Oleg Krichevsky for useful
discussions. A.B.G wishes to thank the Joseph and May Winston
Foundation Career Development Chair in Chemical Engineering, the
Israel Cancer Association (grant No.~20070020B) and the Israel Science
Foundation (grant No.~551/04).


\end{document}